\documentclass[10pt]{article}

\usepackage[OE]{express}
\usepackage{amsmath}
\usepackage{braket}
\usepackage[euler]{textgreek}
\usepackage{graphicx,color,caption}

\begin{document}

\title{Broadband integrated beam splitter using spatial adiabatic passage}

\author{T. Lunghi,\authormark{1*} F. Doutre,\authormark{1} A. P. Rambu,\authormark{2} M. Bellec,\authormark{1} \mbox{M. P. De Micheli,\authormark{1}} A. M. Apetrei,\authormark{2} O. Alibart,\authormark{1} N. Belabas,\authormark{3} S. Tascu,\authormark{2} and S. Tanzilli\authormark{1}}

\address{\authormark{1}Universit\'e C\^{o}te d'Azur, Institut de Physique de Nice (INPHYNI), CNRS UMR 7010, Parc Valrose, 06108 Nice Cedex 2, France.\\
\authormark{2}Research Center on Advanced Materials and Technologies, Sciences Department, Alexandru Ioan Cuza University of Iasi, Blvd. Carol I, nr. 11, 700506 Iasi, Romania.\\
\authormark{3}Centre de Nanosciences et de Nanotechnologies, CNRS UMR 9001, Univ. Paris-Sud, Universit\'e Paris-Saclay, C2N-Marcoussis, 91460 Marcoussis, France.}

\email{\authormark{*}tlunghi@unice.fr}


\begin{abstract*}
Light routing and manipulation are important aspects of integrated optics. They essentially rely on beam splitters which are at the heart of interferometric setups and active routing. The most common implementations of beam splitters suffer either from strong dispersive response (directional couplers) or tight fabrication tolerances (multimode interference couplers). In this paper we fabricate a robust and simple broadband integrated beam splitter based on lithium niobate with a splitting ratio achromatic over more than 130\,nm. Our architecture is based on spatial adiabatic passage, a technique originally used to transfer entirely an optical beam from a waveguide to another one that has been shown to be remarkably robust against fabrication imperfections and wavelength dispersion. Our device shows a splitting ratio of 0.52$\pm $0.03 and 0.48$\pm $0.03 from 1500\,nm up to 1630\,nm. Furthermore, we show that suitable design enables the splitting in output beams with relative phase 0 or $\pi$. Thanks to their independence to material dispersion, these devices represent simple, elementary components to create achromatic and versatile photonic circuits.
\end{abstract*}
%
%
\bibliographystyle{osajnl}
\bibliography{biblio2}

\section{Introduction}
The ability to coherently control the spatial degree of freedom of optical beams is fundamental in integrated optics. This is commonly obtained through beam splitters, optical components able to separate an input signal into two outputs with a specific ratio. 
Directional couplers, the most common architecture to realize integrated beam splitters, suffer from strong dispersive responses. Although chromatic-dependent behaviour may be suitable for certain applications, it is also crucial to realize devices with spectrally flat response~\cite{Adar92}. For instance, achromatic splitters are required in integrated Mach-Zehnder modulators to equally split the optical power between the two paths of the interferometer consistently over a broad spectral range . 

Spatial adiabatic passage (SAP)~\cite{Menchon-Enrich2016} currently emerges as a novel approach to fully transfer an optical beam from a waveguide to another one~\cite{Longhi06}. This mechanism has shown a remarkable robustness against fabrication imperfections and wavelength dispersion. It is also material independent and SAP transfers have been realized on laser written photonic circuits~\cite{Dreisow09}, reconfigurable light-induced waveguide structure~\cite{Ciret2013}, silicon on insulator substrates~\cite{Menchon-Enrich2013} and more recently on lithium niobate substrates~\cite{Chung15, Liu16}. 

In 2009 Dreisow \textit{et al.} exploited an incomplete adiabatic transfer to route a fraction of an optical beam and to realize an achromatic beam splitter~\cite{Dreisow09}. The splitting ratio essentially depended on the geometrical parameters of the structure and weakly depended on the wavelength. This device, referred to as fractional-SAP (f-SAP), featured robust beam splitting power between 500\,nm and 1000\,nm. Later, Chung \textit{et al.}~\cite{Chung12} propose an alternative architecture based on 2-folded-SAPs and showed numerically an extremely broad achromatic response. 
Both these works focused only on the amplitude splitting capabilities but we show here that SAP design can be adapted to engineer 50/50 beam splitters with a phase relationships between the output beams being either 0 or $\pi$ all over the full bandwidth.

In this paper, we experimentally demonstrate a broadband beam splitter using a telecom-compliant coupler integrated on lithium niobate. Our design is based on Chung's proposal~\cite{Chung12}. We demonstrate achromatic splitting power between 1500 and 1630\,nm. Then, we compare a 2-folded-SAP with the f-SAP at telecom wavelengths: both structures feature similar \mbox{amplitude-splitting} capabilities, but the phase relationship between the two output beams is different. 2-folded-SAP provides two output beams in phase while f-SAP provides two beams of opposite phase independently of the wavelength.
These interesting properties demonstrate that these devices represent two complementary building blocks to engineer spatial states of light exploiting SAP.%
\section{Spatial adiabatic passage in theory}
Let us consider an array of straight waveguides along the propagation direction $z$. The propagation of light in the array of coupled waveguides can be modeled using standard coupled-mode theory (CMT) where we consider only coupling between nearest neighbors. 

\begin{figure}[!htbp]
    \centering
    \includegraphics[width = 0.9\textwidth]{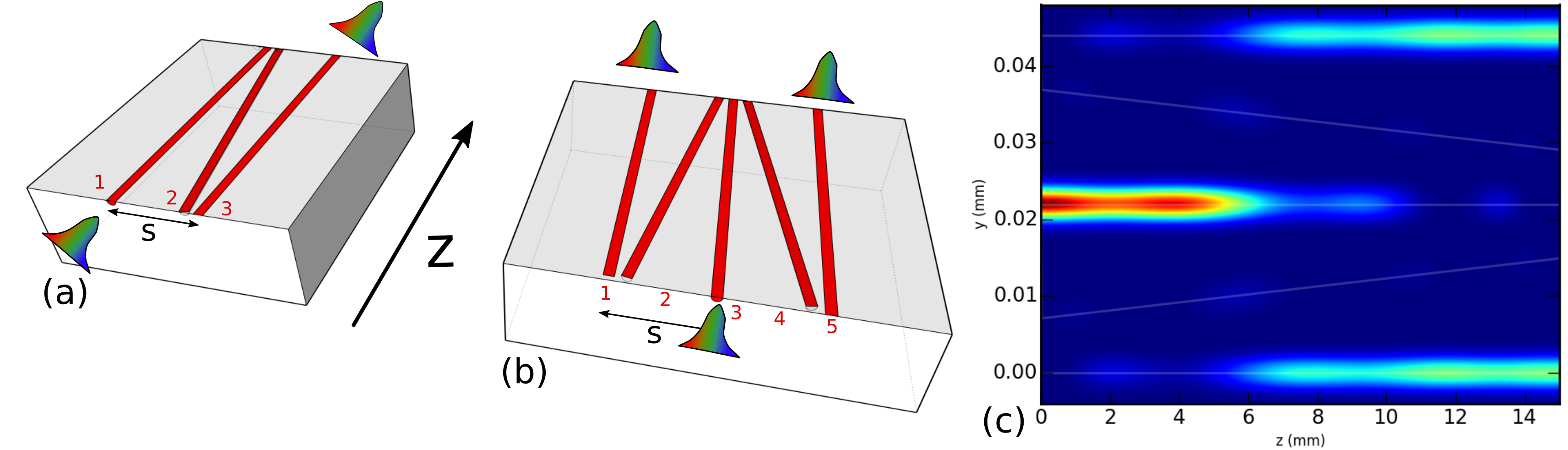}
    \caption{(a) Typical layout of the 3-waveguide SAP. (b) Typical layout of the 2-folded-SAP. (c) Simulation of the evolution of the intensity field along the propagation distance for a 2-folded SAP at 1540\,nm. The dimensions correspond to the experimental values in this work.}
    \label{fig:SAP}
\end{figure}

The typical layout of a 3-waveguide SAP is sketched in Fig.~\ref{fig:SAP}(a)~\cite{Chung15}: a waveguide inclined with an angle $\alpha$ with respect to $z$ direction surrounded by two outer waveguides (1 and 3) parallel to $z$ direction and separated by a distance $s$.
The length of the entire structure is $2L$ with the three waveguides equally spaced at $z = L$.

The spatial evolution of the wave amplitude in each waveguide of the array can be solved using a set of 3 coupled differential equations (see Eq.~\ref{Eq:SAP})~\cite{Paspalakis2006}:

\begin{equation}\label{Eq:SAP}
        -i\frac{d}{dz}
			\begin{pmatrix}
			a_{1}\\
			a_{2}\\
			a_{3}
			\end{pmatrix}
			=
			\begin{pmatrix}
			0	&\kappa_{12}	&	0\\
			\kappa_{21}	&\Delta	&	\kappa_{23}\\
			0	&\kappa_{23}	&	0
			\end{pmatrix}
			\begin{pmatrix}
			a_{1}\\
			a_{2}\\
			a_{3}
			\end{pmatrix}
\end{equation}

where $a_{i}$ is the amplitude of the field in the waveguide $i$, $\Delta=\beta_2-\beta_1=\beta_2-\beta_3$, is the difference of the propagation constants between the waveguide 2 and waveguide 1 (or 3), and $\kappa_{ij} = \kappa_{ji} = \kappa_{ij}(z, \lambda)$ is the coupling between waveguide $i$ and $j$ which depends on $z$ and the beam wavelength. 
This characteristic propagation matrix has a \emph{zero} eigenvalue whose corresponding eigenstate, called \emph{dark state}, yields
\begin{equation}
\ket{\psi_{dark, 3}(z)} \sim  \frac{1}{\sqrt{1+\Theta(z)^2}}\ket{1}-\frac{\Theta(z)}{\sqrt{1+\Theta(z)^2}}\ket{3},
\end{equation}
with $\Theta(z) = \kappa_{12}(z)/\kappa_{23}(z)$. 
Interestingly, the dark state varies with $z$: by opportunely designing the structure so that at $z = 0$, $\kappa_{12}\ll \kappa_{23}$, one can make the dark state coincide with the eigenstate of waveguide 1, i.~e. $\ket{\psi_{dark, 3}(0)}\sim \ket{1}$. Conversely, at $z = 2L$, $\kappa_{12}\gg \kappa_{23}$, so $ \ket{\psi_{dark, 3}(2L)}\sim -\ket{3}$.

Since supermodes are mutually orthogonal, light injected in the dark state remains in this supermode provided the variations are smooth, i.~e. \emph{adiabatic}~\cite{Vitanov1998}. In practice, this corresponds to a perfect transfer of the optical beam from waveguide 1 to waveguide 3. Remarkably this mechanism is intrinsically robust and, in integrated optics, it guarantees robustness against material dispersions and fabrication imperfections. 

The appealing robustness of this structure has stimulated the interests to go beyond the full transfer from waveguide 1 to waveguide 3. For example, by ending the structure at $z = L$ a coherent superposition between the two optical modes is generated~\cite{Dreisow09}, 
\begin{equation}
\ket{\psi_{dark, 3}(L)} \sim  1/\sqrt{2}\ket{1}-1/\sqrt{2}\ket{3}\label{eq:fSTIRAP}
\end{equation}
This structure has been used to realize an achromatic beam splitter~\cite{Dreisow09} at visible wavelengths. It has also been shown to be the analogue implementation of a fractional-STIRAP~\cite{Vewiger2007}, a mechanism used in atomic physics to create superposition of atomic ensembles. Eventually, the phase difference between the output beams is equal to $\pi$ independently of the wavelength.

In the following we present a design alternative to f-SAP to create an optical power beam splitter with different phase relationship between the output beams. The proposed  layout, a simplified version of~\cite{Chung12}, is shown in Fig.~\ref{fig:SAP}(b): two symmetrically folded, 3-waveguide couplers sharing the same central waveguide, 3. Following the same formalism, the spatial evolution equation of this 5-waveguide device correspond to: 
\begin{equation}\label{Eq:MRAP}
		-i\frac{d}{dz}
		\begin{pmatrix}
			a_{1}\\
			a_{2}\\
			a_{3}\\
			a_{4}\\
			a_{5}
		\end{pmatrix}
		=
		\begin{pmatrix}
			0	&\kappa_{12}	&	0&	0&	0\\
			\kappa_{21}	&\Delta	&	\kappa_{23}&	0&	0\\
			0	&\kappa_{23}	&	0&	\kappa_{23}&	0\\
			0	&0&	\kappa_{23}&	\Delta	&	\kappa_{12}\\
			0	&0&	0&	\kappa_{12}&	0
		\end{pmatrix}
		\begin{pmatrix}
			a_{1}\\
			a_{2}\\
			a_{3}\\
			a_{4}\\
			a_{5}
		\end{pmatrix}.
\end{equation}

Note we apply the conditions $\kappa_{23} =\kappa_{34}$ and  $\kappa_{12} =\kappa_{45}$. 
In analogy with the 3-waveguide coupler, the dark state corresponds to 

\begin{equation}\label{eq:MRAP}
\ket{\psi_{dark, 5}(z)} \sim  -\frac{\Theta(z)}{\sqrt{1+\Theta(z)^2}}\ket{1}+\frac{1}{\sqrt{1+\Theta(z)^2}}\ket{3} -\frac{\Theta(z)}{\sqrt{1+\Theta(z)^2}}\ket{5},
\end{equation}

and it evolves adiabatically from $\ket{\psi_{dark, 5}(0)} \sim \ket{3}$ to \mbox{$\ket{\psi_{dark, 5}(2L)} \sim -(\ket{1} +\ket{5})$}. A simulation of the intensity propagation of $\ket{\psi_{dark, 5}(z)}$ is shown in Fig.~\ref{fig:SAP}(c).
If the length of the adiabatic coupler is 2L, this structure acts as a robust, achromatic, beam splitter, similarly to the f-SAP. Notably, the phase difference between the two output modes for the \mbox{f-SAP} and the 2-folded-SAP is $0$ or $\pi$ independently of the wavelength. 

\section{2-folded-SAP design and fabrication}
We optimized the 2-folded-SAP to be implemented on a lithium niobate integrated chip, LiNbO$_3$. The optimal design results from the trade-off between theoretical and practical requirements. On one hand, SAP requires that $\kappa_{12}(0)=\kappa_{23}(2L)\ll \kappa_{23}(0)=\kappa_{12}(2L)$ and \emph{smooth} $z$-variations. On the other hand, in order to be implemented with other photonic structures, 2-folded-SAP design needs to be compact. Larger separations $s$ and angles $\alpha$ allow for more robust devices at the expense of increasing the length of the couplers and (respectively) degrading the adiabaticity conditions. 

The waveguide coupler is designed to work in the telecommunication band around 1.55~\textmu m. Waveguides are fabricated through High-Vacuum Proton-Exchange technique~\cite{Rambu18} (HiVacPE) starting from z-cut substrates. This technique, similar to Soft Proton Exchange (SPE)~\cite{Chanvillard2000}, allows increasing the extraordinary refractive index through the replacement of lithium ions with protons.Compared to other proton-exchange techniques\cite{Bortz1991}, both HiVacPE and SPE ensure controlled proton-substitution rate in order to preserve the crystalline structure of the substrate as well as the nonlinear properties of the material. 

Among the geometric parameters investigated, the best working point in terms of cross talk, broadband operations and compactness, has been found for $2L$~=~1.5~cm, waveguide width equal to 6~\textmu m, s~=~22~\textmu m, and $\alpha$~=~$0.03^{\circ}$. 
The chosen settings ensure $ \kappa_{12}(0) \approx 0.15 \kappa_{23}(0)$, higher than those reported in the literature~\cite{Liu16} but they allow to realize adiabatic passage with structures as short as 1.5~cm. Finally, more compact architectures might be realized by introducing tapered waveguides~\cite{Milton75, Liu16}.

\begin{figure}[!htbp]
\centering
\includegraphics[width = 0.7\textwidth]{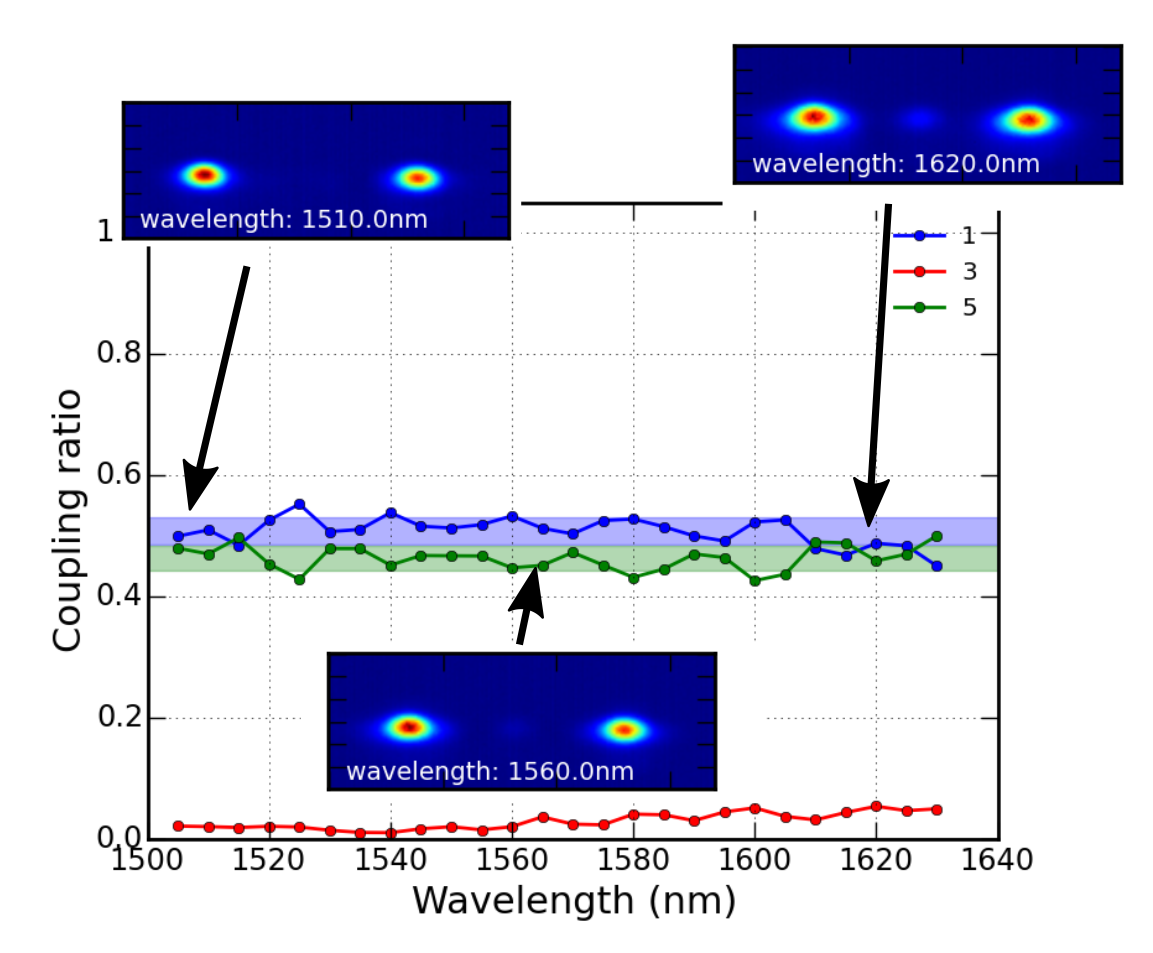}
\caption{Splitting power of 2-folded-SAP as a function of the wavelength. In the insets the images of the output beams at three different wavelengths are reported. Blue (green) area corresponds to the average splitting ratio $0.508\pm 0.022$ ($0.463\pm0.020$)}
\label{fig:output_MRAP}
\end{figure}

The fabricated sample is characterized with a tunable laser diode between 1500 and 1630~nm. Linearly polarized light is coupled in an optical fiber then into the central waveguide. To characterize the splitting ratio we collect the output mode profiles with a 11~mm lens and image them on an InGaAs camera whose linearity in this regime has been characterized. The coupling ratio in waveguide $i$ is measured by calculating the ratio between the output power in the $i-$waveguide and the total power ($P_i/(\sum_i P_i)$). The output power in the $i-$waveguide is calculated by integrating the power in each pixel covering the ith-waveguide. The background noise of the camera is sampled in the not-illuminated areas and then removed. Figure~\ref{fig:output_MRAP} shows the ratio of the output power in waveguide 1, 3 and 5 as a function of the wavelength. The optical power is equally split between the external waveguides and almost no light comes out from the central one. The splitting ratio in waveguide 1 (waveguide 5) is $0.508\pm 0.022$ ($0.463\pm0.020$) averaged over all the measured range and is very close to the ideal 50/50. 
Furthermore no clear drop is measured suggesting that the device has an even broader bandwidth. The quality of the adiabatic transfer is assessed by measuring the residual power exiting in the central waveguide 3 and it is quantified using the cross talk definition $CR = log_{10}(P_3/(\sum_i P_i))$. Cross-talk is smaller than $-15.0\pm 0.2$\,dB in the investigated range. In other words, after having corrected for transmission losses ($\sim$0.4 dB/cm), coupling efficiency (50\%) and reflection at the output facet (84\%), more than 97\% of the optical power is correctly transferred to the output ports. By neglecting the contribution of the central waveguide, the average splitting ratio can be renormalized to $0.52 \pm 0.03$ and $0.48\pm 0.03$.
Asymmetry in the splitting ratio is probably originated by asymmetric losses between the two symmetrical structures. At the extremities of the inclined waveguide, separations between structures are at the limit of our photolithographic equipment ($\sim$1 \textmu m). 

\section{Comparison between 2-folded-SAP and a f-SAP: coherence properties}
We investigate the phase relationship between the output optical beams at the end of the waveguide transfer. In order to fully appreciate the features of this device we compare a 2-folded SAP with a f-SAP power splitter. According to Eq.~\ref{eq:MRAP} and Eq.~\ref{eq:fSTIRAP} both designs result in broadband beam splitting, however phase relationship between the two output beams is $\pi$ for the f-SAP and $0$ for the 5-waveguide power splitter. In order to fabricate the f-SAP we diced and polished a 3-waveguide SAP at $z = L$. Given the limited precision of the dicing/polishing techniques the structure length was smaller than $L$ so the power divider was about 40/60. Better precisions can be obtained using suitable techniques (photolithography, laser written waveguides..) as it has been demonstrated in~\cite{Dreisow09}. Nevertheless this allows us to appreciate the coherence properties of the two structures. 

We perform a classical interference measurement between the output beams by imaging in the far field with the camera: by moving sufficiently far the collection lens, the two output beams interfere as it occurs in a double-slit experiment. If both the beams are in phase, the interference pattern shows a maxima in the central fringe. Conversely, when the two output beams have opposite phases, a dark  interference pattern occurs.

Figure~\ref{fig:coherence_fSTIRAP}(a) and (b) show the interference pattern for the two structures. Since the output beams of a f-SAP have opposite phases (see $\ket{\psi_{dark, 3}(L)}$), the center line is dark. On the contrary, the output beams of the 2-folded-SAP share the same phase therefore the interference pattern results in a maxima in the center line. This result clearly shows the different phase relationship created by the two splitters. Similar patterns have been observed at the different investigated wavelengths, however the limited resolution of this measurement does not allow to assess precise quantitative analysis. This result clearly shows that the two splitters allow obtaining different phase relationships between the two outputs and that this phase relation is stable all over the bandwidth. Knowing that not only the intensity ratio but also the phase relationship is constant all over the bandwidth can find application in originating particular states of light to be used in quantum signal processing.
\begin{minipage}[c]{.50\columnwidth}
    \centering
    \label{fig:coherence_fSTIRAP}\includegraphics[width = .7\textwidth]{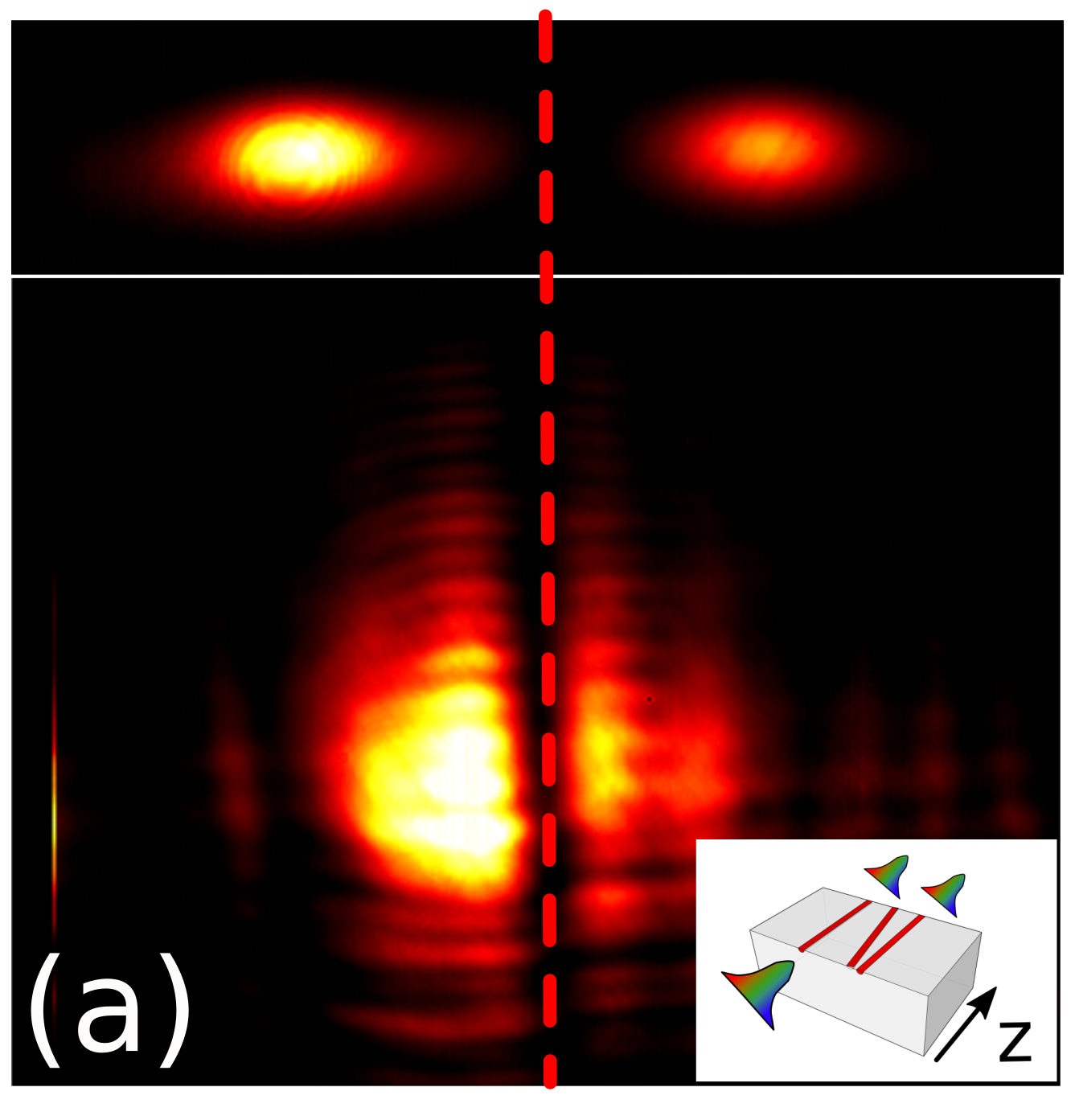}
  \end{minipage}%
  \begin{minipage}[c]{.50\columnwidth}
    \centering
    \includegraphics[width = .7\textwidth]{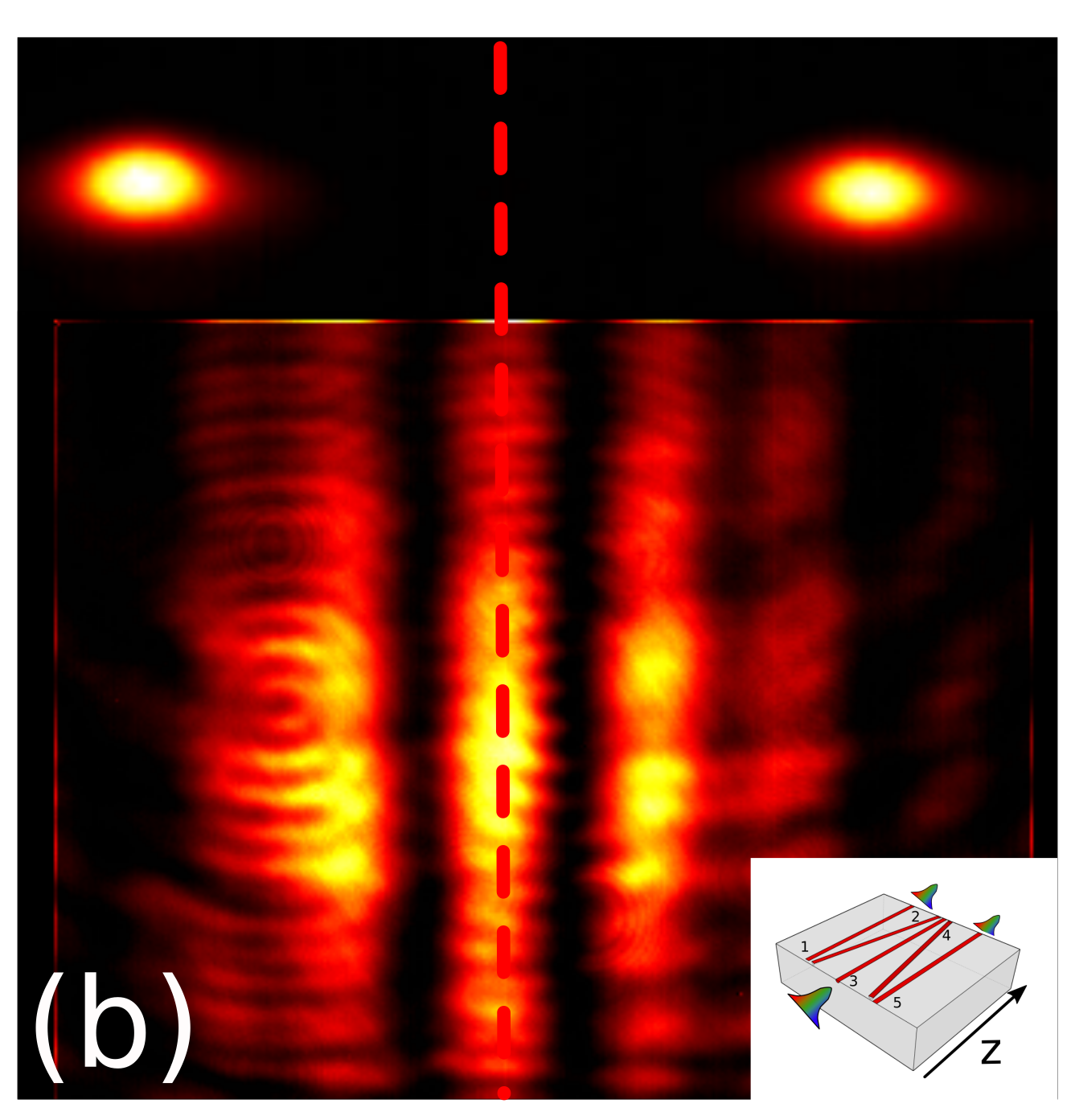}
    \label{fig:coherence_MRAP} 
  \end{minipage}
  \captionof{figure}{far-field interferogram at 1560 nm for f-SAP and 2-folded-SAP. On top an image of the output beams is reported. (a) f-SAP (b) 2-folded-SAP.}
\section{Conclusion}
In this work, we proposed a 2-folded SAP to realize a broadband beam splitter that provides two output beams in phase independently of the wavelength dispersions. We implemented this integrated photonic chip using telecom HiVacPE waveguides on LiNbO$_3$ substrates. We experimentally verified broadband splitting power between 1500 and 1630\,nm. In this range, more than 97\% of the optical power is correctly transferred to the output ports of the splitter. 
We compared this power splitter with fractional-SAP, a similar design proposed in the literature to realize achromatic power splitters. 
While both f-SAP and 2-folded-SAP have similar achromatic splitting capabilities, we proved theoretically and experimentally different phase relationships between the output beams: the 2-folded-SAP creates two output beams in phase while the f-SAP creates two beams out of phase independently of the wavelength.
These two devices represent two complementary building blocks for on-chip engineering of spatial states of light and the achromatic characteristics can expand the capabilities of current integrated optic components. 
\section*{Funding}
Authors acknowledge financial support from the Agence Nationale de la Recherche and from the ``Executive Agency for Higher Education, Research, Development and Innovation Funding'' (UEFISCDI) through the French-Romanian project INQCA (grant agreement PN-II-ID-JRP-RO-FR-2014-0013), This work was conducted within the framework of the project OPTIMAL granted through the European Regional Development Fund (Fond Europeen de developpement regional, FEDER).

\section*{Acknowledgments}
The authors would like to thank H. Tronche, P. Charlier, and G. Schifani, for their help and for fruitful discussions.

\end{document}